\documentclass{article}

\usepackage[english]{babel}

\usepackage[letterpaper,top=2cm,bottom=2cm,left=3cm,right=3cm,marginparwidth=1.75cm]{geometry}

\usepackage{amsmath}
\usepackage{graphicx}
\usepackage{caption}
\usepackage{subcaption}
\usepackage{soul}

\usepackage[colorlinks=true, allcolors=blue]{hyperref}

\begin{document}
{\Large\textbf\newline{Rethinking the external globus pallidus and information flow in cortico-basal ganglia-thalamic circuits}}
\newline \\
Cristina Giossi\textsuperscript{1,2},
Jonathan E. Rubin\textsuperscript{3,4},
Aryn Gittis \textsuperscript{4,5},
Timothy Verstynen\textsuperscript{4,6,$\dagger$} and Catalina Vich\textsuperscript{1,2,$\dagger$}
\newline \\
\textbf{1} Departament de Matemàtiques i Informàtica, Universitat de les Illes Balears, Palma, Spain. \\
\textbf{2} Institute of Applied Computing and Community Code, Palma, Spain.\\
\textbf{3} Department of Mathematics, University of Pittsburgh, Pittsburgh, Pennsylvania, United States of America. \\
\textbf{4} Center for the Neural Basis of Cognition, Pittsburgh, Carnegie Mellon University, Pennsylvania, United States of America. \\
\textbf{5} Department of Biology, Carnegie Mellon University, Pittsburgh, Pennsylvania, United States of America. \\
\textbf{6} Department of Psychology \& Neuroscience Institute, Carnegie Mellon University, Pittsburgh, Pennsylvania, United States of America. \\

\noindent $^\dagger$ {Equally contributing authors.}

\begin{abstract}
\noindent For decades the external globus pallidus (GPe) has been viewed as a passive way-station in the indirect pathway of the cortico-basal ganglia-thalamic (CBGT) circuit, sandwiched between striatal inputs and basal ganglia outputs.
According to this model, one-way descending striatal signals in the indirect pathway amplify the suppression of downstream thalamic nuclei by inhibiting GPe activity.
Here we revisit this assumption, in light of new and emerging work on the cellular complexity, connectivity, and functional role of the GPe in behavior. We show how, according to this new circuit-level logic, the GPe is ideally positioned for relaying ascending and descending control signals within the basal ganglia. Focusing on the problem of inhibitory control, we illustrate how this bidirectional flow of information allows for the integration of reactive and proactive control mechanisms during action selection. Taken together, this new evidence points to the GPe as being a central hub in the CBGT circuit, participating in bidirectional information flow and linking multifaceted control signals to regulate behavior.
\end{abstract}

\section{Introduction}
\label{sec.Introduction}

Imagine walking up to a busy intersection, intending to cross the street. You look at the crosswalk sign and it indicates that it is not safe for you to enter the intersection, so you stop at the corner. The light turns green and the crosswalk sign indicates that it is now safe to proceed. You begin to step off the curb and into the intersection, but you sense something approaching quickly from your side. You stop and step back to the curb before you are even consciously aware of why you have stopped. Then a car shoots in front of you, running the red light and narrowly missing you.  


This scenario highlights some of the multifaceted control processes that regulate our actions \cite{dunovan2015competing, meyer2016neural, aron2011reactive, braver2012variable}. The first stop on the approach to the corner reflects what is known as \textit{proactive} inhibitory control. This form of control represents the cessation of action as an internally generated choice \cite{dunovan2015competing} and is often referred to as a ``no go'' response. The second stop, reflecting an unconscious reaction to the approaching car, illustrates \textit{reactive} inhibitory control \cite{mallet2016arkypallidal, wessel2018surprise}, sometimes referred to as ``braking'' or ``stopping''. This is a faster, and less deliberative, form of inhibition that involves the termination of an ongoing or almost initiated motor plan in response to an external stimulus. 

Both proactive and reactive inhibitory control appear to rely on common neural substrates known as the cortico-basal ganglia-thalamic (CBGT) pathways \cite{nambu2002functional}. These distributed circuits are thought to function as loops that relay information from the cortex to subcortical pathways and back up to the same cortical areas, regulating the tone of cortical activity \cite{mink1996basal,haber2003}. Yet despite this reliance on a common circuit, the means of control for proactive and reactive inhibition 
have been thought to be largely independent within the CBGT network. 
In this view, proactive control relies primarily on the \textit{indirect pathway}, which regulates inhibition of the thalamus via striatopallidal connections. Essentially the indirect pathway, which results in reduced CBGT feedback to cortex, has to ``win'' a competition against the \textit{direct pathway} that runs as a parallel loop and pushes to increase thalamocortical excitatory drive \cite{dunovan2016believer}. In contrast, reactive control has been thought to rely on the so-called \textit{hyperdirect pathway}, bypassing the striatum and regulating inhibition of the thalamus via cortical input to the subthalamic nucleus and its excitatory impact on pallidal nuclei \cite{nambu2002functional}. This model of control via the CBGT pathways relies on two critical assumptions: 1) information flows one-way through the CBGT circuits, and 2) the major pathways in these circuits do not interact with each other before the output stage of processing. 

Recent discoveries over the past 15 years have called these two assumptions into question, however. In particular, new discoveries concerning one critical nucleus in the CBGT circuit, the external segment of the globus pallidus (GPe), reveal a rich intricacy of cell types and connections that are forcing us to reconsider how information flows through the CBGT network \cite{mallet2012dichotomous, dodson2015distinct, abecassis2020npas1+, ketzef2021differential}. 
Here we review the old and new literature on the GPe and its role in behavior, focusing on the process of inhibitory control.
We begin by reviewing the classical stop signal task and models of reactive inhibition via the hyperdirect pathway. We then review recent discoveries about the complexity of the GPe, including novel cell types and connectivity patterns, as well as functional observations, both physiological and behavioral. From a circuit-level logic perspective, we show how these new discoveries point to the GPe as a central hub that relays ascending and descending control signals through the CBGT circuit, linking proactive and reactive mechanisms together.
We finish by highlighting future directions on which the field can focus, to flesh out the nature of the interaction of these pathways and their consequences for behavior and cognition. 

\section{Reactive stopping and the classical model} 

\subsection{Reactive inhibition and the stop signal task} \label{sec.Reactive_inh_SST}

To illustrate the nature of information flow in the CBGT circuit, we will focus on the process of inhibitory control, particularly reactive inhibition. 
We make this choice for two reasons.
First, reactive inhibition is one of the most well-studied paradigms in the context of CBGT circuits and cognition. Second, the classical model of how CBGT circuits implement reactive stopping relies on the two fundamental assumptions of CBGT circuit computation mentioned in Section~\ref{sec.Introduction}: unidirectional information flow and independent control pathways. 

As illustrated in our opening example by the ability to quickly withdraw from a crosswalk as a car approaches, reactive inhibition involves terminating an action or planned action in response to an external stimulus. The most popular paradigm for studying reactive inhibitory control is the classic stop signal reaction time task \cite{vince1948intermittency, lappin1966use}. A systematic review of the task and its limitations is beyond this review's scope (we suggest \cite{verbruggen2019consensus} for any interested readers). Here we provide a short overview of the general stop signal task paradigm in order to walk the reader through the CBGT computations involved.

\begin{figure}
     \centering
     \includegraphics[scale=0.33]{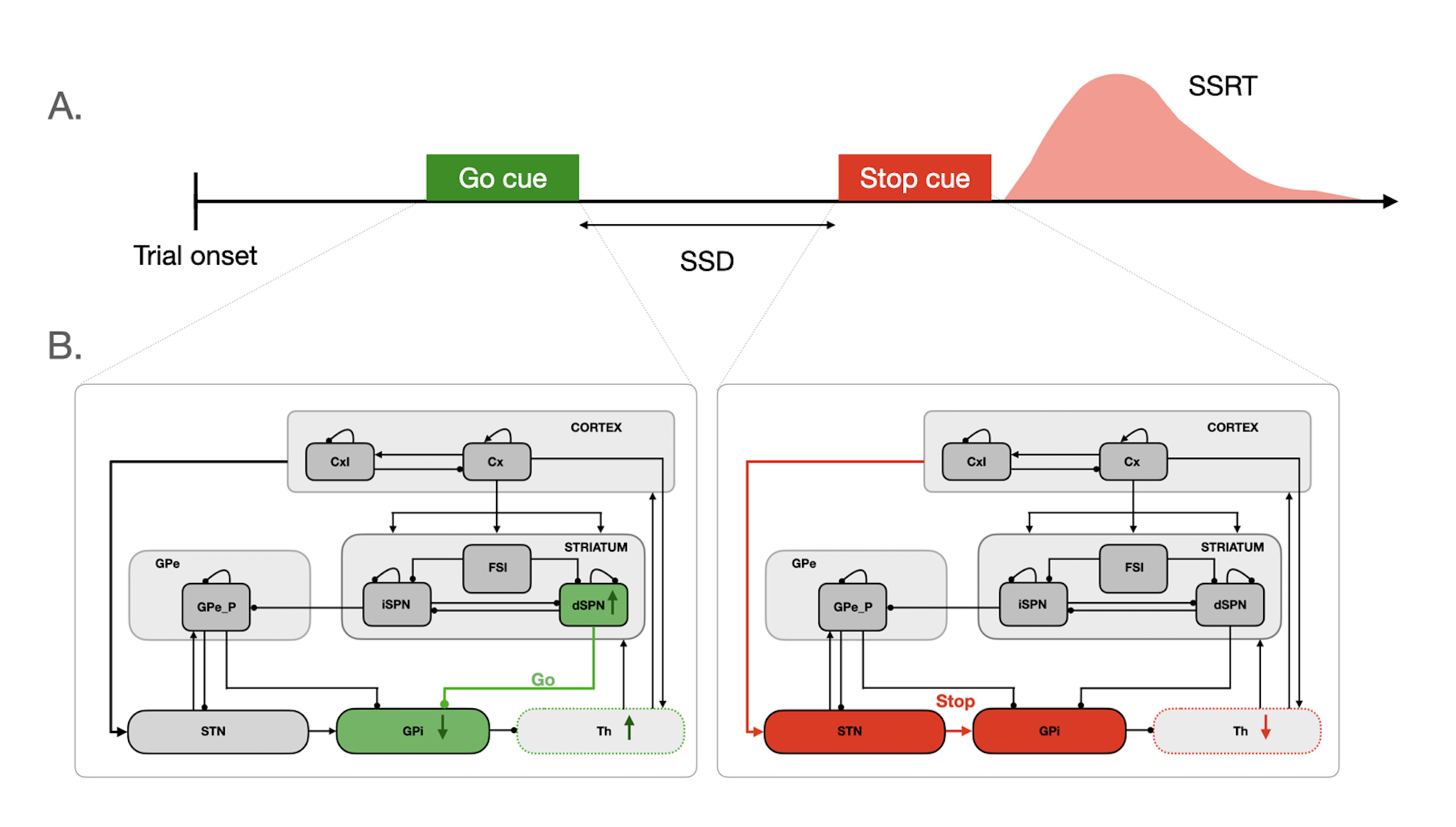} 
     \caption{\textbf{Classical model of reactive inhibition.} A. Schematics of the general stop signal task paradigm. After the trial onset, a primary stimulus (the ``Go" cue, in green) is presented to participants, signaling that they should quickly respond. When a secondary stop stimulus follows (``Stop" cue, in red), participants are expected to withhold their response. The delay between the presentation of the two stimuli is known as stop signal delay (SSD). A useful measure that reflects the time participants take to ``react" to the ``Stop" cue is the stop signal reaction time (SSRT). 
     B. Network schematics showing the dynamics that take place within the CBGT circuit when ``Go" and ``Stop" cues are presented, respectively, according to the classical model. The presentation of the ``Go" cue triggers the activation of the direct pathway (green regions), causing the disinhibition of the thalamus. In the classical view, the ``Stop" cue triggers the engagement of the hyperdirect pathway (red regions), increasing the suppression of thalamic activity.}
     \label{fig:fig1}
\end{figure}

In a typical stop signal task (Figure \ref{fig:fig1}A), participants are asked to quickly respond to a primary stimulus (the ``Go" cue; e.g., pressing a button when they see a green circle) but to withhold that response when they encounter a secondary ``stop" signal (the ``Stop" cue; e.g., a brief tone) that is presented a short time, usually a few hundred milliseconds, after the primary stimulus. The timeline between the ``Go" cue and the ``Stop" signal, known as the stop signal delay (SSD), is usually varied - either dynamically adapted based on the participant's performance or sampled at specific intervals. 
The participant's ability to stop when experiencing the different SSDs is used to estimate their stop signal reaction time (SSRT). The SSRT reflects the median time it takes to ``react" to the stop signal and successfully inhibit the action. 

The stop signal task has become a popular tool due to its sensitivity at detecting individual differences across a range of subject groups. In clinical populations, elevated SSRTs (indicating poorer inhibitory control) have been associated with various pathologies such as attention-deficit/hyperactivity disorder \cite{alderson2007attention}, substance abuse disorders \cite{fillmore2002impaired} and obsessive-compulsive disorder \cite{chamberlain2006motor}. Such findings suggest that compromised reactive inhibitory control may play a role in the etiology or maintenance of these disorders. In non-clinical populations, SSRTs have been linked to individual differences in personality traits like impulsivity \cite{logan1997impulsivity} and have been used to investigate cognitive changes across the lifespan \cite{williams1999development}, with SSRTs appearing to initially shorten during early development and then lengthen with aging after adulthood. The popularity of the stop signal task, as well as its sensitivity at spotting individual differences in both clinical and non-clinical populations, makes it an ideal paradigm for describing the process of reactive inhibitory control. We now move on to consider traditional models for how the stop signal task is implemented in CBGT pathways.

\subsection{Classical model: Unidirectional flow} \label{sec.Classic_model}

The classical model of CBGT inhibitory control in the stop signal task posits a one-way flow of information ``downward" in the CBGT circuit, with proactive (striatal-initiated) and reactive (STN-initiated) control arising from independent sources. 
According to this model \cite{aron2006cortical}, the ``Go" stimulus triggers engagement of the \textit{direct pathway}, starting with glutamatergic, excitatory signals 
from cortical regions to D1-expressing spiny projection neurons (SPNs) in the striatum. The direct pathway SPNs (dSPNs) then send GABAergic inhibition into the main output nucleus of the basal ganglia, which in primates is the internal globus pallidus (GPi). The GPi has tonically active neurons that send GABAergic, inhibitory projections into the matrix of the thalamus \cite{kita2005balance, nambu2000excitatory}. Thus, inhibiting the GPi increases the activity of the thalamus through disinhibition, potentially amplifying cortical activity via thalamocortical pathways and increasing the likelihood of a response. 

Running parallel to the direct pathway is the \textit{indirect pathway}. This pathway is also driven by excitatory signals from the cortex, but in this case, they terminate on D2-expressing SPNs. These indirect pathway SPNs (iSPNs) send GABAergic projections to the GPe. 
The GPe cells, in turn, inhibit both the subthalamic nucleus (STN) and the GPi, forming the so-called long and short indirect pathways, respectively. The net result of the engagement of both branches of the indirect pathway is the increased activity of GPi cells, which strengthens the suppression of their thalamic targets and reduces activity in the thalamocortical loops. These effects decrease the likelihood of a subsequent response. 

A third canonical control pathway, known as the \textit{hyperdirect pathway} \cite{nambu2002functional}, runs through the CBGT circuit. This more recently discovered circuit bypasses the striatum altogether, by sending glutamatergic projections to the STN itself. Because the STN directly projects to GPi, this architecture provides a rapid, two synapses link from the cortex to the outputs of the basal ganglia, proposed as a faster control mechanism than the indirect pathway that can implement an urgent command to interrupt evolving action plans.

In the classical model of reactive stopping, only the hyperdirect pathway is thought to regulate reactive control \cite{nambu2002functional,aron2006cortical}. Following the timeline of the stop signal task (Figure~\ref{fig:fig1}A), this model proposes that the ``Go" cue initiates activation of the direct pathway (Figure~\ref{fig:fig1}B, left panel), thereby starting a drive process that, if left unchecked, will eventually lead to the triggering of a response via disinhibition of the thalamus. The presentation of the ``Stop" cue activates the hyperdirect pathway (Figure~\ref{fig:fig1}B, right panel), which quickly boosts GPi activity and stop the action initiation process. If the hyperdirect pathway can sufficiently rapidly achieve and maintain the suppression of the thalamus,  
then a successful reactive stop occurs. Otherwise, the direct pathway succeeds, and an erroneous go response is produced.

The classical model of reactive control arose mainly from a combination of inferring function from the logic of the canonical CBGT circuit as understood at the time (see next section for more on this) \cite{aron2016frontosubthalamic}, as well as correlational evidence of fast STN activation in response to a stop signal (e.g., \cite{aron2006cortical,chen2020prefrontal,sano2013signals, wadsley2022stopping}). Baked into this model are some critical assumptions. First, it assumed that the diffuse excitatory projections from STN to GPi are strong enough to prevent a significant reduction  of GPi activity due to direct pathway inhibition. This assumption leads to the second assumption: information only flows down to the GPi within CBGT circuits. This unidirectional model of information flow means that reactive control, implemented via hyperdirect pathway signals, does not interact with more proactive control mechanisms, which involve a competition between the direct and indirect pathways. As we will soon show, these two assumptions have come into question as our understanding of the CBGT circuit, and particularly our knowledge about the GPe nucleus, has expanded in recent years. New discoveries have forced the field to rethink the circuit-level architecture of the CBGT pathways and how information flows through this circuit to contribute to behavioral control. 


\section{Emerging views of the GPe} \label{sec.GPe_emreging_views}

While the subject of a recent surge in attention, aspects of the cellular complexity of the GPe, including the identification of prototypical and arkypallidal neuron subtypes, have been known for over 50 years \cite{delong1971activity}. Yet the functions of the various cell types in the GPe and their roles in guiding behavior were, until recently, largely overlooked. The GPe was essentially treated as a homogenous node in the CBGT circuit, with primarily a descending influence on information flow from the STN to the GPi. This simplified view changed with the advent of new methods that allowed researchers to study how GPe cells can be classified from a molecular perspective, as well as in terms of their electrophysiological properties, axonal projections, or dendritic morphology. Here we summarize the current understanding of the cellular composition, connectivity, and functional properties of this nucleus. This rundown allows us to lay out the foundation for rethinking the role of the GPe in CBGT circuit computation and behavioral control. For a more complete summary of the anatomy and physiology of the GPe we point the interested reader to \cite{dong2021connectivity} and \cite{courtney2023cell}. 

\subsection{Cellular composition} \label{sec.CellComposition}

The GPe is mostly comprised of GABAergic neurons, with only about 5\% of its cells being cholinergic \cite{abdi2015prototypic, hernandez2015parvalbumin+, abecassis2020npas1+, mastro2014transgenic} (Figure \ref{fig:fig2}A). Typically researchers focus on the non-cholinergic cells, characterizing the GPe in terms of two principal classes of inhibitory neurons: almost 70\% of its neurons are labeled as \textit{prototypical} (GPe\_P in Figure~\ref{fig:fig2}), whereas \textit{arkypallidal} neurons (GPe\_A in Figure~\ref{fig:fig2}) represent approximately 20\% of the neurons in the GPe. Prototypical cells themselves form a heterogeneous class of neurons. In some studies, they are labeled based on the expression of a specific calcium-binding protein, parvalbumin (PV), as a molecular marker \cite{mallet2012dichotomous, abdi2015prototypic}. Others prefer to cluster prototypical cells based on the expression of transcription factors such as Nkx2.1 (NK2 homeobox 1) \cite{dodson2015distinct} and Lhx6 (LIM homeobox 6) \cite{abdi2015prototypic}. On the other hand, arkypallidal neurons present a unique molecular signature, expressing the opioid precursor preproenkephalin (PPE) and the forkhead box protein P2 (FoxP2) \cite{mallet2012dichotomous, abdi2015prototypic, dodson2015distinct}. In some studies, Npas1 (neuronal PAS domain protein 1), another protein-coding gene, is used to label arkypallidal neurons, since it largely overlaps with FoxP2-expressing neurons \cite{pamukcu2020parvalbumin+}.

One key property that has been used to distinguish between prototypical and arkypallidal neurons is their firing rates. This distinction was first observed in Delong's original analysis of cells in non-human primates \cite{delong1971activity}. He found that in dopamine-intact \textit{in vivo} conditions, prototypical neurons had reliable spontaneous firing rates ranging from $1$ to $100$ spikes/s, with an average of $50-55$ spike/s overall (see also \cite{dodson2015distinct}). On the other hand, arkypallidal neurons had more irregular and sporadic activity in both \textit{in vivo} and \textit{ex vivo} conditions, with overall lower firing rates ranging from $1$ to $30$ spike/s ($10$ spikes/s average). This firing dropped during sleep, unlike prototypical neurons (see also \cite{mallet2012dichotomous, gittis2014new, abdi2015prototypic, dodson2015distinct, mallet2016arkypallidal, pamukcu2020parvalbumin+, aristieta2021disynaptic, ketzef2021differential}). These observations highlight how these two cell populations have clearly distinct spike rate characteristics that distinguish them, along with their different underlying molecular signatures.

There is also some evidence to suggest that the cells in these two GPe neuron classes have somewhat different morphologies \cite{mallet2012dichotomous}. 
Arkypallidal neurons' axons appear to be characterized by lengths far exceeding those of prototypical neurons and dendritic spines with significantly higher density. On the other hand, prototypical cells have been shown to feature significantly longer local axon collaterals and a larger number of synaptic boutons compared to arkypallidals. More work is still needed to confirm these morphological distinctions between the two major cell types, however.
Despite these morphological distinctions, prototypical and arkypallidal neurons appear to be intermingled throughout the GPe \cite{delong1971activity, dodson2015distinct}, and 
this 
homogeneous distribution appears to be relatively consistent across the rostral, central, and caudal segments of the GPe \cite{abdi2015prototypic}.

Collectively, the prototypical and arkypallidal cells account for about 95\% of all GABAergic GPe neurons \cite{abdi2015prototypic}.  Given their molecular and electrophysiological distinctions, it seems reasonable to suspect that they contribute differently to GPe's role in regulating the information flow through the CBGT circuits, which raises the possibility of a more sophisticated role for the GPe than the classical theories suggest. 

\subsection{Connectivity} \label{sec.Connectivity}

\begin{figure}
     \centering
     \includegraphics[scale=0.45]{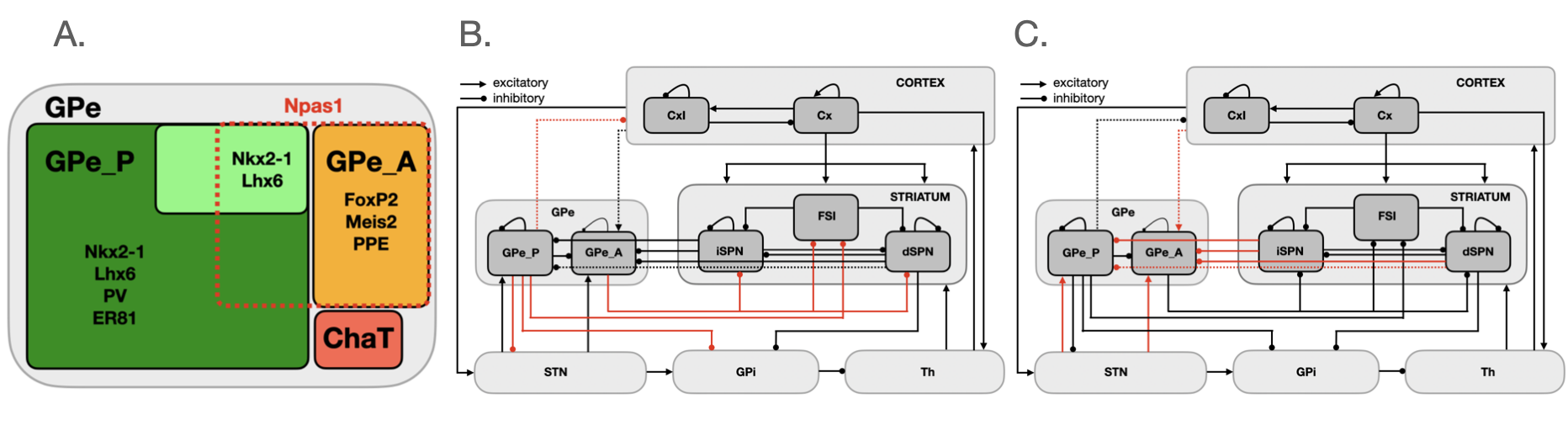} 
     \caption{\textbf{Cellular composition and connectivity of the GPe.} A. Schematic diagram illustrating the cellular composition of the GPe. The areas of the rectangles represent the approximate sizes of the corresponding neuron classes. Prototypical neurons, which constitute $\sim$70\% of GPe inhibitory cells, can be subdivided according to whether or not they are PV-expressing. Arkypallidal neurons represent $\sim$20\% of GPe GABAergic neurons; all of them and a small pool of prototypical neurons are Npas1-expressing. ChAT-expressing neurons constitute $\sim$5\% of the total GPe neuron population, showing no overlap with other known clusters of GPe cells. B. Network diagram of  the CBGT circuit, highlighting (in red) the efferent projections of GPe. C. Network diagram of the CBGT circuit, highlighting (in red) the afferent projections to GPe.}
     \label{fig:fig2}
\end{figure}

Another way to categorize GPe neurons
is by the nature of their connections. 
In this view, two different groups of neurons emerge: one relays information downward within the basal ganglia, while the other participates in the flow of information upwards through the basal ganglia. It is generally accepted that the former group aligns with the prototypical neurons, while the latter are mostly arkypallidal neurons. 
The efferent and afferent pathways from these two cell types are shown in Figure \ref{fig:fig2}B and \ref{fig:fig2}C, respectively. Here we provide a brief overview of the established afferent, efferent, and collateral projections of the GPe. 

We begin with the outward, efferent projections from the GPe. The only cell class known to make synapses with the STN and the GPi (or substantia nigra pars reticulata, SNr) are the prototypical neurons \cite{abdi2015prototypic, mallet2012dichotomous, mastro2014transgenic, saunders2016globus, glajch2016npas1+, aristieta2021disynaptic} (Figure \ref{fig:fig2}B).  Arkypallidal  neurons instead provide GABAergic innervation - around a thousand axonal boutons per arkypallidal neuron \cite{mallet2012dichotomous, fujiyama2016single, dong2021connectivity} - across the striatum, projecting to direct and indirect pathway SPNs as well as to striatal fast spiking interneurons (FSIs) \cite{mallet2012dichotomous, gittis2014new, abdi2015prototypic, hernandez2015parvalbumin+, saunders2015direct, fujiyama2016single, corbit2016pallidostriatal, pamukcu2020parvalbumin+, aristieta2021disynaptic, ketzef2021differential}. Still little is known about the relative strength of arkypallidal projections to striatum. Glajch et al \cite{glajch2016npas1+}, however, assessed that arkypallidal neurons project more strongly to iSPN neurons than to dSPNs, with a 2:1 ratio.
Prototypical neurons also have been determined to project to striatum, although experimental findings do not agree on the target populations, how dense these projections are, or to what extent they are functionally relevant. According to some studies, prototypical neurons have been shown to have dense projections into the striatum \cite{fujiyama2016single}, while other studies suggest that the density of striatal connections from prototypical cells may be more modest \cite{mallet2012dichotomous}. These projections appear to mainly target the FSI cells in the striatum \cite{gittis2014new, abdi2015prototypic, saunders2016globus, glajch2016npas1+}, and with more strength than that with which arkypallidal neurons signal to FSIs \cite{corbit2016pallidostriatal}.
Interestingly, findings have revealed the existence of GPe projections to the neocortex itself \cite{saunders2015direct, chen2015identification}, particularly to cortical motor areas from a specific subset of GPe neurons co-expressing Npas1 and Nkx2.1 and from GPe cholinergic neurons \cite{abecassis2020npas1+}, suggesting that the GPe could be involved in the regulation of premotor and motor activity through a cortico-pallidal-cortical loop \cite{chen2015identification, courtney2023cell}. It remains to be determined whether there are other pallidal cell subtypes that also project to the neocortex.

As for the afferent projections, most of the existing evidence points to distinct input connections to prototypical and arkypallidal neurons (Figure \ref{fig:fig2}C). Indeed, both the STN and striatum have been shown to differentially innervate GPe cells. STN and iSPNs provide more robust inputs to prototypicals than to arkypallidal neurons \cite{pamukcu2020parvalbumin+, aristieta2021disynaptic, gast2021role, ketzef2021differential}. In particular, iSPN projections to arkypallidal neurons have been estimated to be 85\% weaker than those targeting prototypical neurons and also less numerous \cite{aristieta2021disynaptic}, while STN inputs have been measured to be 74\% weaker to arkypallidal than prototypical cells \cite{aristieta2021disynaptic}. 
In addition, emerging studies have reported the existence of other GABAergic projections coming from the striatum, specifically from dSPN neurons and preferentially targeting arkypallidal neurons \cite{ketzef2021differential, cui2021striatal} or, more broadly, Npas1 neurons \cite{labouesse2023non}. For consistency's sake, we will refer to these as arkypallidal neurons, even though not all Npas1 neurons are of this type. Given the nature of the suppressive, outward efferent projections from GPe\_A neurons (Figure~\ref{fig:fig2}B), these inhibitory inputs from dSPNs may promote action initiation via inhibition of arkypallidal cells, constituting a so-called non-canonical striatopallidal ``Go" pathway \cite{ketzef2021differential, aristieta2021disynaptic, labouesse2023non}. Little, however, is known about the specifics of this connection. It could be possible, indeed, that the inhibition promoted by dSPNs onto arkypallidal neurons is topographically organized into functional units that encode specific motor patterns and outcome behaviors \cite{labouesse2023non}. 
Moreover, Aristieta et al. (2021) showed that \textit{in vivo} opto-stimulation of iSPN neurons disinhibits arkypallidal neurons through a disynaptic circuit, suggesting that dSPN-pallidal projections could balance arkypallidal outputs through an indirect local competition with iSPN-GPe inputs \cite{labouesse2023non}.
These ideas about dSPN impact on GPe\_A, however, contrast with other findings that dSPNs exclusively affect prototypical neurons \cite{mizutani2017substance}. More work is required to resolve this discrepancy. Lastly, cortical projections have been identified as the source of almost 10\% of the total input into GPe, preferentially targeting arkypallidal neurons \cite{abecassis2020npas1+, karube2019motor}. Indeed, experiments suggest that only one third to half of prototypical neurons receive cortical inputs \cite{abecassis2020npas1+}.

Finally, we turn to the internal connectivity of the GPe. Within the GPe, strong collateral GABAergic projections from prototypical to arkypallidal neurons have been observed and are thought to play a fundamental role in switching the activity of arkypallidal neurons on and off, according to whether prototypical neurons are inhibited or not, respectively \cite{mallet2012dichotomous, dodson2015distinct, fujiyama2016single, aristieta2021disynaptic, ketzef2021differential}. What is less certain is whether these collateral connections are reciprocal. Optogenetic excitation of arkypallidal neurons has not been shown to produce inhibitory responses in prototypical neurons, in either \textit{ex vivo} or \textit{in vivo} \cite{aristieta2021disynaptic} conditions, suggesting either that synapses from arkypallidal onto prototypical neurons do not exist or that their influence is fairly weak \cite{mallet2012dichotomous, ketzef2021differential, gast2021role}. Lastly, some studies suggest that prototypical neurons exhibit a stronger degree of intra-population inhibition than do arkypallidal neurons \cite{ketzef2021differential, gast2021role}. In contrast, others  have assessed that the strengths of connections between arkypallidal neurons seem modest, but connections between prototypical cells are even weaker \cite{nevado2014effective}.

\subsection{Functional roles of GPe neuron subtypes in behavior} \label{sec.GPerole}

As was pointed out in Section~\ref{sec.Classic_model}, the GPe was traditionally considered to be a node of the ``motor-suppressing" indirect pathway, conveying descending signals to the output nuclei of the basal ganglia circuit \cite{calabresi2014direct}. Indeed, according to the classical model, striatal iSPN neurons send the majority of their projections to the GPe, which then exerts an inhibitory influence on the STN and the GPi \cite{kita2007globus}. 
Taking into account the excitatory connections from STN to GPe immediately complicates this picture. As a specific example, if GPe activity increases and results in strengthened inhibition to the STN, then the GPe neurons lose some of their excitatory input and hence may reduce their activity back towards baseline levels.  This feedback loop has long been recognized as a possible source of parkinsonian oscillations when dopaminergic effects become compromised \cite{plenz1999,terman2002}.
More recently, the emerging work of the past decade has cast doubt on the simple, one-dimensional view of the functional role of the GPe \cite{mallet2012dichotomous, abdi2015prototypic, dodson2015distinct, abecassis2020npas1+}. With the discovery of new cell types within the GPe and new connections (e.g., arkypallidal projections to striatum; Section~\ref{sec.Connectivity}), a renewed interest has emerged in understanding the functional properties of this nucleus in light of its complex connectivity and cellular composition. 

It is now known that the GPe contributes to both motor and non-motor functions and communicates with other basal ganglia nuclei and other brain regions through both upstream and downstream projections (Figure~\ref{fig:fig2}B-C). Amongst the non-motor roles of GPe cells, a small pool of PV-expressing GPe neurons seems to be associated with reversal learning and processing of sensory and reward cues \cite{courtney2023cell, lilascharoen2021divergent, farries2023selective}. Consistent with this, a recent study by Isett et al. (2023) showed that inhibition of PV cells in the GPe drove transient punishment of behavior, not motor suppression \cite{isett2023indirect}, highlighting the role of GPe in learning processes, as well as movement control. Evidence suggests that other GPe cells, likely Npas1-Nkx2.1-expressing, could be involved in regulating sleep \cite{qiu2016deep, vetrivelan2010role, lazarus2013role} and limbic functions \cite{stephenson2016basal, wallace2017genetically}. Dysfunctions of GPe contribute to several clinical conditions such as Parkinson's disease \cite{mallet2008parkinsonian, gittis2014new, crompe2020globus, courtney2023cell}, Huntington’s disease \cite{courtney2023cell, starr2008pallidal, beste2015behavioral, deng2021progression} and dystonia \cite{baron2011multi, nambu2011reduced, starr2005spontaneous, chiken2008cortically}. Analysis of the various functional roles of the GPe would be worthy of an entire review in and of itself (see also \cite{courtney2023cell, dong2021connectivity, hegeman2016external}). For simplicity's sake, we will focus here on those roles related to action control.

There is now a consistent body of evidence linking variations in firing rate of GPe neurons with the dynamics of movement \cite{yoshida2016two, gu2020globus, arkadir2004independent}, as well as a variety of more direct causal tests of their role in motor control \cite{glajch2016npas1+, pamukcu2020parvalbumin+, aristieta2021disynaptic, cui2021dissociable, mastro2017cell, lilascharoen2021divergent}. Indeed, GPe neuron firing patterns are not only correlated with movement specifics, such as amplitude, velocity, and direction \cite{georgopoulos1983relations, mitchell1987primate, gage2010selective}, but have also been found to tune for the body region involved, and the nature of the movement itself (e.g., whether is passive, active, or externally cued) \cite{georgopoulos1983relations, turner1997pallidal, turner2005context, gage2010selective}.
In this vein, it is important to recognize that even though GPe output is inhibitory, it is nonetheless possible that it passes along informative signals to its synaptic targets via deviations from its baseline, pacemaker-like firing (see e.g. \cite{corbit2016pallidostriatal} for analogous effects related to inhibitory outputs of FSIs).  

Within the context of the various cell types in the GPe, it has been proposed that there is a cell-type-specific encoding of spontaneous movement in the GPe \cite{dodson2015distinct}. According to this model, the prototypical neurons show heterogeneous firing responses, while arkypallidal neurons present robust increases in their firing profile during spontaneous movements. Here, the decrease in firing of GPe cells, particularly prototypical neurons, from activation of striatal neurons during movement \cite{cui2013concurrent} would reflect the traditional role of the GPe as an arm of the indirect pathway \cite{dodson2015distinct, kravitz2010regulation, sano2013signals}. In contrast, since arkypallidal neurons show little firing at rest \cite{mallet2012dichotomous, dodson2015distinct, delong1971activity} and robustly increase activity around movement onset \cite{dodson2015distinct}, they could be engaged in action facilitation. Inhibiting large striatal regions could prevent competing actions from being expressed in order to promote the selection of a desired action \cite{mallet2012dichotomous, aristieta2021disynaptic, ketzef2021differential, glajch2016npas1+, hegeman2016external}. For the remainder of this section, we will focus on this distinction in the roles of prototypical and arkypallidal neurons in the action selection (or inhibition) process.


Much of the recent work examining the role of the GPe cell types in motor control and action selection has focused primarily on the process of inhibitory control, with most of the emphasis placed on the contribution of arkypallidal neurons \cite{mallet2016arkypallidal, schmidt2017pause, dodson2015distinct}. Mallet et al. (2016) \cite{mallet2016arkypallidal} examined the activity of the two GPe subpopulations in a stop signal task scenario (Section~\ref{sec.Reactive_inh_SST}) and found that the time courses of both prototypical and arkypallidal firing exhibit a clear increase following the presentation of a stop signal. Arkypallidal neurons produce a significantly stronger response to the stop cue than prototypical neurons do, suggesting that arkypallidals have a greater influence on the production of a stop response. This enhancement of arkypallidal activity occurs just before the surge of movement-related striatal activity, as would be expected if the arkypallidal neurons play an important role in canceling imminent actions. A more recent study by Aristieta et al. (2021) \cite{aristieta2021disynaptic} supported this hypothesis by using \textit{in vivo} optogenetic stimulation of arkypallidal FoxP2-expressing neurons during the same stop signal task (see Figure 7E-I from \cite{aristieta2021disynaptic}). This manipulation produced a strong inhibition of ongoing locomotion, providing causal evidence that activation of arkypallidal neurons is sufficient to induce this effect, likely through a global suppression of go-related striatal activity. Similarly, Pamucku et al. (2020) \cite{pamukcu2020parvalbumin+} showed that optogenetic stimulation of Npas1-expressing neurons, a subset of cells within the GPe mostly consisting of arkypallidal neurons but also including a small fraction of prototypical neurons \cite{abdi2015prototypic}, induces a decrease in the vigor of motor output, both in terms of duration and speed of movement (see Figure 3B-bottom panel from \cite{pamukcu2020parvalbumin+}). To confirm that the movement-suppression effect enhanced by Npas1-expressing neurons upon optogenetic stimulation is mediated through the inhibition of the dorsal striatum, they also optogenetically stimulated axon terminals of Npas1-expressing neurons in the dorsal striatum directly.  This stimulation produced a decrease in locomotion, which provides further support for the hypothesis that the arkypallidal pathways play a pivotal role in movement inhibition.

On the other hand, there is no consensus on the involvement of prototypical cells in reactive inhibition. Some lines of evidence have shown that prototypical neurons also become mildly excited during the presentation of a stop signal \cite{mallet2016arkypallidal}. This evidence, however, must be contrasted with evidence suggesting that prototypical neurons do not play any role in the context of reactive inhibition \cite{aristieta2021disynaptic}. One way to reconcile these disparate observations comes from Aristieta et al. (2021) \cite{aristieta2021disynaptic}, who showed that inhibitory inputs from axon collaterals from prototypical neurons control the activity of arkypallidal neurons. Specifically, the authors found that optogenetic excitation of prototypical neurons produced a strong inhibition of the activity of recorded arkypallidal cells. The influence of prototypical neurons on movement suppression then makes sense given the circuit-level logic of the striato-pallidal-striatal loop. Specifically, activation of iSPNs inhibits prototypical neurons that, in turn, inhibit arypallidal units that inhibit dSPNs. In this way, there is a secondary arm of the canonical indirect pathway that further suppresses the motor-promoting signals from dSPNs through disinhibition of arkypallidal cells. These findings provide evidence for the existence of a wiring architecture between prototypical and arkypallidal cells that is capable of regulating the activity of arkypallidal neurons, suggesting that the prototypical neurons could play an indirect role in the control of reactive inhibition via modulation of arkypallidal activity. Moreover, Gage et al. (2010) suggest that a sharp decrease of GPe prototypical neurons plays a role in information processing during behavioral choice tasks \cite{gage2010selective}. Indeed, the disinhibition provoked by prototypical neurons onto FSIs, causes a coordinated pulse of increasing activity in FSIs as chosen actions are initiated while suppressing unwanted alternatives.

\section{Rethinking the classical model: GPe as a central hub}





 

\begin{figure}[!ht]
    \centering
     \includegraphics[width = 1.0\textwidth]{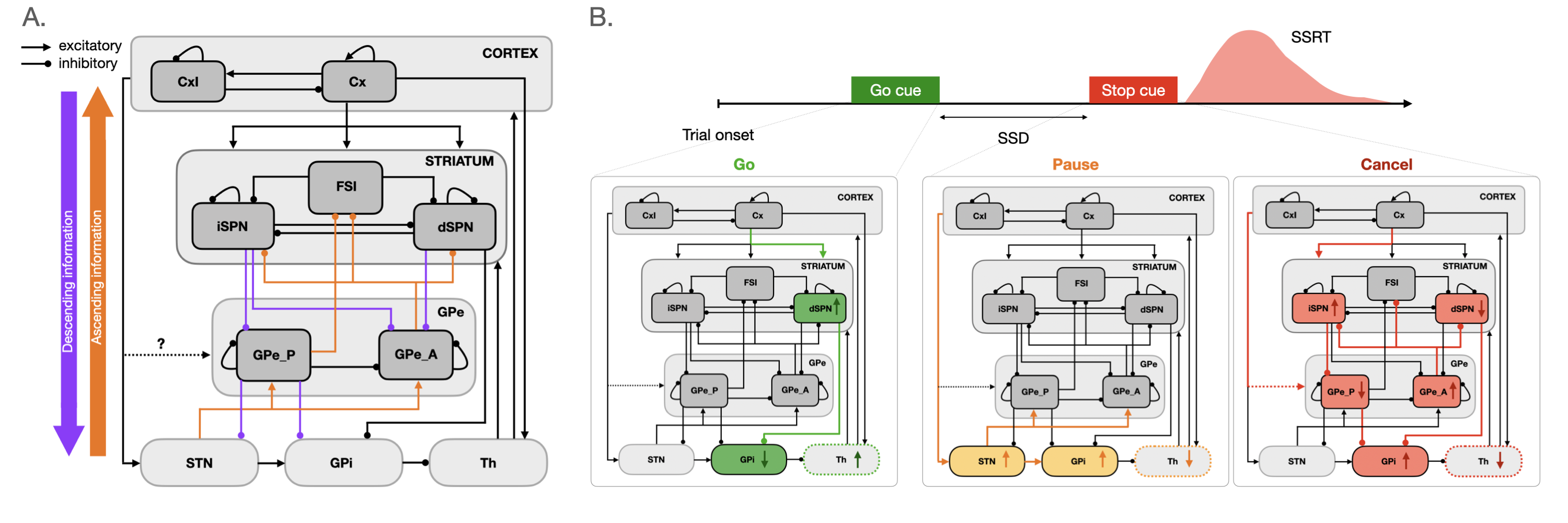}
     \caption{\textbf{Casting the GPe as a central hub of the basal ganglia with a key role in regulating information flow in cortico-basal ganglia-thalamic circuits.} A. CBGT network schematic. Here, the GPe is placed in a central position, regulating the bidirectional information flow across the circuit (descending pathways in purple, ascending pathways in orange). The cortical projection to GPe is indicated as a dotted line with  a question mark because there is still little known about its nature. 
     B. Network dynamics for action suppression according to the Pause-then-Cancel model. As in the classical view, the ``Go" cue triggers the activation of the direct pathway, leading to thalamic disinhibition. In this model, the ``Stop" signal triggers two subprocesses: the first fast process activated is a ``pause" process, induced by the engagement of the hyperdirect pathway (in yellow), which transiently increases the inhibition of the thalamus. The second subprocess corresponds to the ``cancel" stage of the model when arkypallidal neurons become engaged via stop-related signals from the STN or through direct cortical inputs and effectively relay this instruction upwards to the striatum.} 
    \label{fig:fig4}
\end{figure}

These new insights into the organization, connectivity, and functional roles of the GPe fundamentally shift our understanding of this nucleus and its role in regulating the flow of information through CBGT circuits. While the classical model posits the GPe as a nucleus with a homogenous neuronal composition, we now recognize a rich complexity of cell types in the GPe with a qualitative distinction into two major classes: prototypical and arkypallidal neurons. Also in the classical model, the GPe only projects to the STN and GPi/SNr, thus relaying only descending signals to the output of the basal ganglia. We now know, however, that the GPe also sends ascending signals to the striatum, directly onto striatal SPNs and FSIs. Thus, the influence of the GPe goes both up and down relative to the traditional basal ganglia pathways. Finally, the classical model interpreted the GPe as a simple way station along the movement-inhibiting pathway originating from the iSPNs (i.e., indirect pathway). New experimental evidence paints a much more complex picture of the GPe's functional role in the motor domain, with seemingly mixed results about its influence on subsequent behavior. These fundamental changes in our understanding of the GPe suggest that we need to rethink the role that this nucleus plays in CBGT computations. 
The terms descending and ascending information still make sense, because they are defined in terms of the dominant basal ganglia output nuclei, the GPi and SNr, being downstream.  In this new view, however, the GPe is centrally located to regulate bidirectional information flow. The path of descending information (purple connections in Figure~\ref{fig:fig4}A) is an extension of the classical model, with striatal SPNs sending signals to the GPe that are then relayed down to the STN and GPi via prototypical cells.
In addition, ascending information (orange connections, Figure~\ref{fig:fig4}A), originating either from hyperdirect pathway drive to STN or possibly from direct cortical projections to the GPe \cite{karube2019motor}, propagates up to the striatum via arkypallidal GPe neurons and regulates SPN firing through GABAergic signaling. Moreover, the two GPe populations interact, at least through prototypical inhibition of arkypallidal GPe neurons, setting up a possible mechanism for dominance to switch between the two directions.  In this way, information flow through the CBGT circuit is no longer unidirectional and the control of ascending and descending information is centrally regulated by the GPe. 

We can appreciate this central role of the GPe in regulating bidirectional information flow most clearly in the process of reactive inhibition. In particular, the recently proposed \textit{Pause-then-Cancel} model \cite{mallet2016arkypallidal,schmidt2017pause} highlights the complementary roles of ascending and descending signals in reactive stopping (Figure~\ref{fig:fig4}B). Framed in the context of the typical stop signal task (Section~\ref{sec.Reactive_inh_SST}), the Pause-then-Cancel model still separates two competing control signals from cortex: the imperative ``Go" signal from cortex to the striatum, particularly the direct pathway, and the reactive ``Stop" signal from the hyperdirect pathway. When a primary stimulus is presented (the ``Go" cue), cortical inputs drive the activity of dSPNs (direct pathway; green path in Figure~\ref{fig:fig4}B). These then inhibit the GPi, reducing its inhibition on the thalamus and increasing the likelihood of an action. The Pause-then-Cancel model of reactive stopping begins with the same preliminary step as the classical model of reactive inhibition (Section~\ref{sec.Classic_model}), in which the hyperdirect pathway quickly activates the STN, sending a surge of excitation to the GPi and thus increasing its inhibition of the thalamus (yellow path in Figure~\ref{fig:fig4}B). In this model, however, this drive signal is not sufficient to fully cancel the planned action \cite{mallet2016arkypallidal,schmidt2017pause}, but appears only to delay the progression of the ramping up of thalamic excitation that would promote motor behavior. It is at this stage of the process that the Pause-then-Cancel model deviates from the classical model.

Specifically, the ``cancel" stage of this model (red path in Figure~\ref{fig:fig4}B) relies heavily on the bidirectional influence of the GPe in order to fully implement the cancellation of the planned response. Starting with the ascending flow of information, arkypallidal neurons become engaged either via drive signals from the STN or direct cortical input signals \cite{karube2019motor}. Since the sole known efferents of the arkypallidal neurons are inhibitory connections to the striatum, we can imagine that this activity reduces responses in the direct and indirect pathways. This reduction helps to maintain the suppression of thalamic ramping until cortical drive to iSPNs can take over. This begins a stage where descending information flow becomes the dominant factor in the stopping process, which happens in two parts. First, because of the inhibition of dSPNs, GPi is released from inhibition. Second, because iSPNs are being engaged, the balance of power shifts to the traditional indirect pathway. The iSPNs predominantly inhibit prototypical cells in the GPe, which are the only pallidal cells to project downward to the STN (long indirect pathway) and GPi (short indirect pathway). Thus, GPi output can increase, providing an enhanced suppression of the thalamic response and leading to the cessation of the planned response.

The Pause-then-Cancel model articulated within the context of our expanded knowledge of the GPe clearly highlights two critical aspects of CBGT computation. First, the GPe links proactive (direct/indirect pathways) and reactive (hyperdirect pathway) inhibitory mechanisms together. Second, the GPe is ideally situated to regulate the bidirectional flow of information through the CBGT circuit. Indeed, the second feature is a mechanism for the first. Specifically, the GPe plays a key role in relaying ascending reactive control signals to the striatal pathways that implement proactive control through descending projections. Thus, the GPe acts as  
a central hub within the CBGT circuit, rather than being an isolated component of the indirect pathway alone.

\section{Looking forward} 

As we have shown, the emerging evidence on the cellular composition, connectivity, and functional characteristics of the GPe fundamentally changes our understanding of the role that this nucleus plays in regulating information flow through the CBGT circuit. While the classical model considers the GPe as a homogeneous nucleus, we now know that it is composed of a variety of cell types that can be divided into at least two general categories: prototypical and arkypallidal neurons. The rediscovery of a heterogeneous collection of GPe cell types has led to renewed interest in the anatomical and functional characteristics of each subpopulation. Prototypical neurons continue to be seen as an integral part of the classical indirect pathway through which striatal iSPN signals impact the activity downstream in the STN and GPi. Interestingly, a new twist on this idea is that prototypical GPe neuron firing rates may tune the chloride load and hence GABA reversal potential for GPi (or SNr) neurons, thereby impacting how strongly dSPN inputs affect GPi firing rates and yielding a new form of interaction between the direct and indirect pathways \cite{phillips2020}.  On the other hand, arkypallidal neurons regulate ascending information through striatum-targeting projections. In light of this new evidence, we now see that the GPe assumes a central role in basal ganglia computations, becoming not just a relay station of the indirect pathway, but a pivotal hub of the full CBGT circuit, regulating both ascending and descending information streams. Using the Pause-then-Cancel model as an example, we have highlighted how the GPe is ideally situated to integrate signals from different cortical sources in order to implement behavioral control. In particular, this integration arises in stopping, because this process combines both reactive and proactive control signals to cancel planned actions. This example supports the idea that as a field, we should shift our conceptualization of the GPe from being an incidental node along the indirect pathway to a central hub that integrates signals from all three canonical CBGT pathways.



It is worth noting that this view of the central role of the GPe in basal ganglia function has been recently proposed by Courtney et al. (2023) \cite{courtney2023cell}. In this review, the authors go over the same cellular and circuit-level discoveries as we review here, arriving at a similar conclusion that the GPe acts to regulate more distributed aspects of basal ganglia network function than previously thought. Unlike the current perspective piece, however, their work primarily focuses on how the GPe contributes to the etiology of disease states. For example, the loss of dopamine that characterizes Parkinson's disease correlates with alterations in GPe neuronal activity. Indeed, Npas1 neurons show hypoactivity, while the STN input strength to GPe PV neurons is reduced \cite{pamukcu2020parvalbumin+}. Moreover, a late stage in the progression of Huntington's disease causes the loss of arkypallidal neurons \cite{deng2021progression}, while dystonia symptomatology seems to rely on reduced activity of GPe PV neurons due to compromised hyperpolarization and cyclic nucleotide-gated (HCN) channels \cite{chiken2008cortically}. Here we complement this perspective on pathologies of the CBGT circuit by focusing more explicitly on the computational role that the GPe plays in normative function. Putting the two together, the disease states described by Courtney et al. (2023) can be understood in terms of alterations in the bidirectional information processing that occurs in normal basal ganglia dynamics. Nonetheless, this similarity in conclusions highlights the converging new view of the field on the role of the GPe in CBGT circuit dynamics.

Theoretically, this shift in perspective forces a fundamental change in how we think about information flow in CBGT circuits. Traditional models of these pathways describe a one-way architecture, where information conveyed by cortex propagates ``down" towards the output nuclei, the GPi and SNr \cite{mink1996basal, alexander1986parallel}. This has been the dominant view of CBGT information flow for over a half-century, reflecting the ``independent" aspect of the ``parallel and independent" pathways framework \cite{alexander1986parallel}. However, if the GPe regulates ascending information flow, as well as descending information, allowing signals originating at the STN (or GPe itself) to influence striatal computation, then our collective understanding of the circuit needs to be updated. The CBGT pathways in this new view comprise more a complex, recurrent network architecture that allows for the integration of signals from multiple cortical (and possibly subcortical) sources. No pathway in the CBGT circuit can be considered as fully ``independent" anymore, and our models of CBGT computation need to be revised to reflect the interactions involved.


With this new understanding, many new questions arise. As we point out in Section~\ref{sec.CellComposition}, a consensus about the cellular composition of the GPe has yet to be reached. The pool of existing evidence characterizing the GPe cell populations remains small and incomplete. The heterogeneity of the GPe, in fact, goes beyond the simple dichotomous organization of GABAergic neurons into prototypical and arkypallidal cells (see Figure~\ref{fig:fig2}A). Indeed, this represents only a conveniently simplified reduction of the true underlying cellular complexity of the GPe. If different neuron types have different anatomical and functional properties, then the nature of GPe computations is likely still more complex than what we have discussed. 
Even within the dichotomous classification, unknowns remain. For example, are arkypallidal neurons homogenous in terms of their involvement in inhibitory control or is there functional variability across arkypallidal subpopulations? In addition, arkypallidal cells receive afferents from multiple sources, including the striatum, STN, and recently discovered inputs from cortex \cite{abecassis2020npas1+, karube2019motor}. How do arkypallidals integrate these disparate inputs and are there yet unknown afferents that remain to be discovered?
Moreover, recent data suggest the possibility of heterogeneous neural subtypes within target nuclei of the GPe, setting up the possibility of a more complex collection of parallel pathways than those included in the classical basal ganglia model \cite{delgado2023}. We are at the tip of the proverbial iceberg in terms of our understanding of this nucleus, and CBGT circuits more broadly. 


Nevertheless, this new perspective on the CBGT circuit already raises a critical question: what advantages do a bidirectional flow of information, and the increased resource demands that come along with it, provide in terms of CBGT computation? Building off of our focus on inhibitory control, we can ask how ascending projections from arkypallidal neurons influence striatal computation. Do arkypallidal neurons serve to amplify inhibition promoted by the indirect pathway alone or does their role rely more on shifting the balance of power between iSPNs and dSPNs \cite{dunovan2016believer}? The answer to this question has critical implications for the nature of behavioral control mediated by these circuits. 

\section*{Acknowledgements}

The authors would like to thank Jyotika Bahuguna and Nico Mallet for helpful discussions on the GPe that informed the structure of this review.

\section*{Conflict of interest}
The authors have no conflict of interest to report.

\section*{Funding Information}
CG and CV are supported by the PCI2020-112026 project, and CV is also supported by the PCI2023-145982-2, both funded by MCIN/AEI/10.13039/501100011033 and by the European Union ``NextGenerationEU"/PRTR as part of the CRCNS program.
TV and JER are partly supported by NIH awards R01DA053014 and R01DA059993 as part of the CRCNS program. AG and JER are partly supported by NIH award R01NS125814, also part of the CRCNS program.

\section*{Author contributions}
CG and TV conceived the idea of this manuscript, reviewed the literature, and wrote the manuscript. 
JER, CV, and AG provided many useful suggestions and helped to edit this manuscript. All authors read and approved the final manuscript.

\clearpage

\begin{figure}
    \centering
    \includegraphics[scale=0.35]{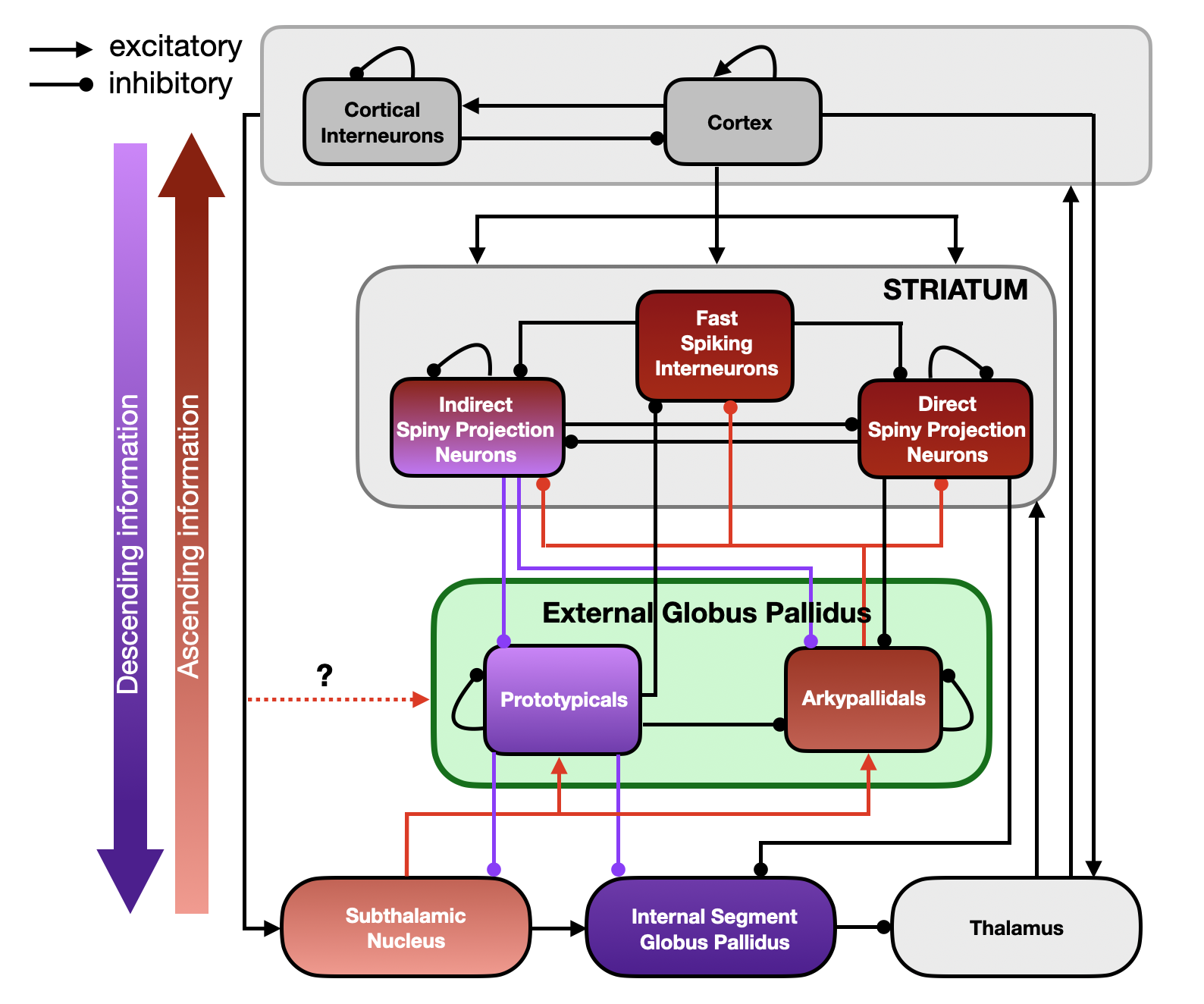} 
    \caption*{\textbf{GRAPHICAL ABSTRACT } 
    The GPe regulates information flow in cortico-basal ganglia-thalamic circuits, relaying ascending and descending signals and serving as a central hub within the CBGT network.}
    \label{fig:fig0}
\end{figure}

\clearpage

\newpage


\begin{thebibliography}{10}
\providecommand{\url}[1]{#1}
\csname url@samestyle\endcsname
\providecommand{\newblock}{\relax}
\providecommand{\bibinfo}[2]{#2}
\providecommand{\BIBentrySTDinterwordspacing}{\spaceskip=0pt\relax}
\providecommand{\BIBentryALTinterwordstretchfactor}{4}
\providecommand{\BIBentryALTinterwordspacing}{\spaceskip=\fontdimen2\font plus
\BIBentryALTinterwordstretchfactor\fontdimen3\font minus
  \fontdimen4\font\relax}
\providecommand{\BIBforeignlanguage}[2]{{%
\expandafter\ifx\csname l@#1\endcsname\relax
\typeout{** WARNING: IEEEtran.bst: No hyphenation pattern has been}%
\typeout{** loaded for the language `#1'. Using the pattern for}%
\typeout{** the default language instead.}%
\else
\language=\csname l@#1\endcsname
\fi
#2}}
\providecommand{\BIBdecl}{\relax}
\BIBdecl

\bibitem{dunovan2015competing}
K.~Dunovan, B.~Lynch, T.~Molesworth, and T.~Verstynen, ``Competing basal
  ganglia pathways determine the difference between stopping and deciding not
  to go,'' \emph{Elife}, vol.~4, p. e08723, 2015.

\bibitem{meyer2016neural}
H.~C. Meyer and D.~J. Bucci, ``Neural and behavioral mechanisms of proactive
  and reactive inhibition,'' \emph{Learning \& Memory}, vol.~23, no.~10, pp.
  504--514, 2016.

\bibitem{aron2011reactive}
A.~R. Aron, ``From reactive to proactive and selective control: developing a
  richer model for stopping inappropriate responses,'' \emph{Biological
  psychiatry}, vol.~69, no.~12, pp. e55--e68, 2011.

\bibitem{braver2012variable}
T.~S. Braver, ``The variable nature of cognitive control: a dual mechanisms
  framework,'' \emph{Trends in cognitive sciences}, vol.~16, no.~2, pp.
  106--113, 2012.

\bibitem{mallet2016arkypallidal}
N.~Mallet, R.~Schmidt, D.~Leventhal, F.~Chen, N.~Amer, T.~Boraud, and J.~D.
  Berke, ``Arkypallidal cells send a stop signal to striatum,'' \emph{Neuron},
  vol.~89, no.~2, pp. 308--316, 2016.

\bibitem{wessel2018surprise}
J.~R. Wessel, ``Surprise: A more realistic framework for studying action
  stopping?'' \emph{Trends in cognitive sciences}, vol.~22, no.~9, pp.
  741--744, 2018.

\bibitem{nambu2002functional}
A.~Nambu, H.~Tokuno, and M.~Takada, ``Functional significance of the
  cortico--subthalamo--pallidal ‘hyperdirect’pathway,'' \emph{Neuroscience
  research}, vol.~43, no.~2, pp. 111--117, 2002.

\bibitem{mink1996basal}
J.~W. Mink, ``The basal ganglia: focused selection and inhibition of competing
  motor programs,'' \emph{Progress in neurobiology}, vol.~50, no.~4, pp.
  381--425, 1996.

\bibitem{haber2003}
S.~N. Haber, ``The primate basal ganglia: parallel and integrative networks,''
  \emph{Journal of Chemical Neuroanatomy}, vol.~26, no.~4, pp. 317--330, 2003.

\bibitem{dunovan2016believer}
K.~Dunovan and T.~Verstynen, ``Believer-skeptic meets actor-critic: rethinking
  the role of basal ganglia pathways during decision-making and reinforcement
  learning,'' \emph{Frontiers in neuroscience}, vol.~10, p. 106, 2016.

\bibitem{mallet2012dichotomous}
N.~Mallet, B.~R. Micklem, P.~Henny, M.~T. Brown, C.~Williams, J.~P. Bolam,
  K.~C. Nakamura, and P.~J. Magill, ``Dichotomous organization of the external
  globus pallidus,'' \emph{Neuron}, vol.~74, no.~6, pp. 1075--1086, 2012.

\bibitem{dodson2015distinct}
P.~D. Dodson, J.~T. Larvin, J.~M. Duffell, F.~N. Garas, N.~M. Doig,
  N.~Kessaris, I.~C. Duguid, R.~Bogacz, S.~J. Butt, and P.~J. Magill,
  ``Distinct developmental origins manifest in the specialized encoding of
  movement by adult neurons of the external globus pallidus,'' \emph{Neuron},
  vol.~86, no.~2, pp. 501--513, 2015.

\bibitem{abecassis2020npas1+}
Z.~A. Abecassis, B.~L. Berceau, P.~H. Win, D.~Garcia, H.~S. Xenias, Q.~Cui,
  A.~Pamukcu, S.~Cherian, V.~M. Hern{\'a}ndez, U.~Chon \emph{et~al.},
  ``Npas1+-nkx2. 1+ neurons are an integral part of the
  cortico-pallido-cortical loop,'' \emph{Journal of Neuroscience}, vol.~40,
  no.~4, pp. 743--768, 2020.

\bibitem{ketzef2021differential}
M.~Ketzef and G.~Silberberg, ``Differential synaptic input to external globus
  pallidus neuronal subpopulations in vivo,'' \emph{Neuron}, vol. 109, no.~3,
  pp. 516--529, 2021.

\bibitem{vince1948intermittency}
M.~A. Vince, ``The intermittency of control movements and the psychological
  refractory period,'' \emph{British Journal of Psychology}, vol.~38, no.~3, p.
  149, 1948.

\bibitem{lappin1966use}
J.~S. Lappin and C.~W. Eriksen, ``Use of a delayed signal to stop a visual
  reaction-time response.'' \emph{Journal of Experimental Psychology}, vol.~72,
  no.~6, p. 805, 1966.

\bibitem{verbruggen2019consensus}
F.~Verbruggen, A.~R. Aron, G.~P. Band, C.~Beste, P.~G. Bissett, A.~T. Brockett,
  J.~W. Brown, S.~R. Chamberlain, C.~D. Chambers, H.~Colonius \emph{et~al.},
  ``A consensus guide to capturing the ability to inhibit actions and impulsive
  behaviors in the stop-signal task,'' \emph{elife}, vol.~8, p. e46323, 2019.

\bibitem{alderson2007attention}
R.~M. Alderson, M.~D. Rapport, and M.~J. Kofler,
  ``Attention-deficit/hyperactivity disorder and behavioral inhibition: a
  meta-analytic review of the stop-signal paradigm,'' \emph{Journal of abnormal
  child psychology}, vol.~35, pp. 745--758, 2007.

\bibitem{fillmore2002impaired}
M.~T. Fillmore and C.~R. Rush, ``Impaired inhibitory control of behavior in
  chronic cocaine users,'' \emph{Drug and alcohol dependence}, vol.~66, no.~3,
  pp. 265--273, 2002.

\bibitem{chamberlain2006motor}
S.~R. Chamberlain, N.~A. Fineberg, A.~D. Blackwell, T.~W. Robbins, and B.~J.
  Sahakian, ``Motor inhibition and cognitive flexibility in
  obsessive-compulsive disorder and trichotillomania,'' \emph{American journal
  of psychiatry}, vol. 163, no.~7, pp. 1282--1284, 2006.

\bibitem{logan1997impulsivity}
G.~D. Logan, R.~J. Schachar, and R.~Tannock, ``Impulsivity and inhibitory
  control,'' \emph{Psychological science}, vol.~8, no.~1, pp. 60--64, 1997.

\bibitem{williams1999development}
B.~R. Williams, J.~S. Ponesse, R.~J. Schachar, G.~D. Logan, and R.~Tannock,
  ``Development of inhibitory control across the life span.''
  \emph{Developmental psychology}, vol.~35, no.~1, p. 205, 1999.

\bibitem{aron2006cortical}
A.~R. Aron and R.~A. Poldrack, ``Cortical and subcortical contributions to stop
  signal response inhibition: role of the subthalamic nucleus,'' \emph{Journal
  of Neuroscience}, vol.~26, no.~9, pp. 2424--2433, 2006.

\bibitem{kita2005balance}
H.~Kita, Y.~Tachibana, A.~Nambu, and S.~Chiken, ``Balance of monosynaptic
  excitatory and disynaptic inhibitory responses of the globus pallidus induced
  after stimulation of the subthalamic nucleus in the monkey,'' \emph{Journal
  of Neuroscience}, vol.~25, no.~38, pp. 8611--8619, 2005.

\bibitem{nambu2000excitatory}
A.~Nambu, H.~Tokuno, I.~Hamada, H.~Kita, M.~Imanishi, T.~Akazawa, Y.~Ikeuchi,
  and N.~Hasegawa, ``Excitatory cortical inputs to pallidal neurons via the
  subthalamic nucleus in the monkey,'' \emph{Journal of neurophysiology},
  vol.~84, no.~1, pp. 289--300, 2000.

\bibitem{aron2016frontosubthalamic}
A.~R. Aron, D.~M. Herz, P.~Brown, B.~U. Forstmann, and K.~Zaghloul,
  ``Frontosubthalamic circuits for control of action and cognition,''
  \emph{Journal of Neuroscience}, vol.~36, no.~45, pp. 11\,489--11\,495, 2016.

\bibitem{chen2020prefrontal}
W.~Chen, C.~de~Hemptinne, A.~M. Miller, M.~Leibbrand, S.~J. Little, D.~A. Lim,
  P.~S. Larson, and P.~A. Starr, ``Prefrontal-subthalamic hyperdirect pathway
  modulates movement inhibition in humans,'' \emph{Neuron}, vol. 106, no.~4,
  pp. 579--588, 2020.

\bibitem{sano2013signals}
H.~Sano, S.~Chiken, T.~Hikida, K.~Kobayashi, and A.~Nambu, ``Signals through
  the striatopallidal indirect pathway stop movements by phasic excitation in
  the substantia nigra,'' \emph{Journal of Neuroscience}, vol.~33, no.~17, pp.
  7583--7594, 2013.

\bibitem{wadsley2022stopping}
C.~G. Wadsley, J.~Cirillo, A.~Nieuwenhuys, and W.~D. Byblow, ``Stopping
  interference in response inhibition: behavioral and neural signatures of
  selective stopping,'' \emph{Journal of Neuroscience}, vol.~42, no.~2, pp.
  156--165, 2022.

\bibitem{delong1971activity}
M.~R. DeLong, ``Activity of pallidal neurons during movement.'' \emph{Journal
  of neurophysiology}, vol.~34, no.~3, pp. 414--427, 1971.

\bibitem{dong2021connectivity}
J.~Dong, S.~Hawes, J.~Wu, W.~Le, and H.~Cai, ``Connectivity and functionality
  of the globus pallidus externa under normal conditions and parkinson's
  disease,'' \emph{Frontiers in neural circuits}, vol.~15, p. 645287, 2021.

\bibitem{courtney2023cell}
C.~D. Courtney, A.~Pamukcu, and C.~S. Chan, ``Cell and circuit complexity of
  the external globus pallidus,'' \emph{Nature Neuroscience}, pp. 1--13, 2023.

\bibitem{abdi2015prototypic}
A.~Abdi, N.~Mallet, F.~Y. Mohamed, A.~Sharott, P.~D. Dodson, K.~C. Nakamura,
  S.~Suri, S.~V. Avery, J.~T. Larvin, F.~N. Garas \emph{et~al.}, ``Prototypic
  and arkypallidal neurons in the dopamine-intact external globus pallidus,''
  \emph{Journal of Neuroscience}, vol.~35, no.~17, pp. 6667--6688, 2015.

\bibitem{hernandez2015parvalbumin+}
V.~M. Hern{\'a}ndez, D.~J. Hegeman, Q.~Cui, D.~A. Kelver, M.~P. Fiske, K.~E.
  Glajch, J.~E. Pitt, T.~Y. Huang, N.~J. Justice, and C.~S. Chan,
  ``Parvalbumin+ neurons and npas1+ neurons are distinct neuron classes in the
  mouse external globus pallidus,'' \emph{Journal of Neuroscience}, vol.~35,
  no.~34, pp. 11\,830--11\,847, 2015.

\bibitem{mastro2014transgenic}
K.~J. Mastro, R.~S. Bouchard, H.~A. Holt, and A.~H. Gittis, ``Transgenic mouse
  lines subdivide external segment of the globus pallidus (gpe) neurons and
  reveal distinct gpe output pathways,'' \emph{Journal of Neuroscience},
  vol.~34, no.~6, pp. 2087--2099, 2014.

\bibitem{pamukcu2020parvalbumin+}
A.~Pamukcu, Q.~Cui, H.~S. Xenias, B.~L. Berceau, E.~C. Augustine, I.~Fan,
  S.~Chalasani, A.~W. Hantman, T.~N. Lerner, S.~M. Boca \emph{et~al.},
  ``Parvalbumin+ and npas1+ pallidal neurons have distinct circuit topology and
  function,'' \emph{Journal of Neuroscience}, vol.~40, no.~41, pp. 7855--7876,
  2020.

\bibitem{gittis2014new}
A.~H. Gittis, J.~D. Berke, M.~D. Bevan, C.~S. Chan, N.~Mallet, M.~M. Morrow,
  and R.~Schmidt, ``New roles for the external globus pallidus in basal ganglia
  circuits and behavior,'' \emph{Journal of Neuroscience}, vol.~34, no.~46, pp.
  15\,178--15\,183, 2014.

\bibitem{aristieta2021disynaptic}
A.~Aristieta, M.~Barresi, S.~A. Lindi, G.~Barriere, G.~Courtand,
  B.~de~La~Crompe, L.~Guilhemsang, S.~Gauthier, S.~Fioramonti, J.~Baufreton
  \emph{et~al.}, ``A disynaptic circuit in the globus pallidus controls
  locomotion inhibition,'' \emph{Current Biology}, vol.~31, no.~4, pp.
  707--721, 2021.

\bibitem{saunders2016globus}
A.~Saunders, K.~W. Huang, and B.~L. Sabatini, ``Globus pallidus externus
  neurons expressing parvalbumin interconnect the subthalamic nucleus and
  striatal interneurons,'' \emph{PloS one}, vol.~11, no.~2, p. e0149798, 2016.

\bibitem{glajch2016npas1+}
K.~E. Glajch, D.~A. Kelver, D.~J. Hegeman, Q.~Cui, H.~S. Xenias, E.~C.
  Augustine, V.~M. Hern{\'a}ndez, N.~Verma, T.~Y. Huang, M.~Luo \emph{et~al.},
  ``Npas1+ pallidal neurons target striatal projection neurons,'' \emph{Journal
  of Neuroscience}, vol.~36, no.~20, pp. 5472--5488, 2016.

\bibitem{fujiyama2016single}
F.~Fujiyama, T.~Nakano, W.~Matsuda, T.~Furuta, J.~Udagawa, and T.~Kaneko, ``A
  single-neuron tracing study of arkypallidal and prototypic neurons in healthy
  rats,'' \emph{Brain Structure and Function}, vol. 221, pp. 4733--4740, 2016.

\bibitem{saunders2015direct}
A.~Saunders, I.~A. Oldenburg, V.~K. Berezovskii, C.~A. Johnson, N.~D. Kingery,
  H.~L. Elliott, T.~Xie, C.~R. Gerfen, and B.~L. Sabatini, ``A direct gabaergic
  output from the basal ganglia to frontal cortex,'' \emph{Nature}, vol. 521,
  no. 7550, pp. 85--89, 2015.

\bibitem{corbit2016pallidostriatal}
V.~L. Corbit, T.~C. Whalen, K.~T. Zitelli, S.~Y. Crilly, J.~E. Rubin, and A.~H.
  Gittis, ``Pallidostriatal projections promote $\beta$ oscillations in a
  dopamine-depleted biophysical network model,'' \emph{Journal of
  Neuroscience}, vol.~36, no.~20, pp. 5556--5571, 2016.

\bibitem{chen2015identification}
M.~C. Chen, L.~Ferrari, M.~D. Sacchet, L.~C. Foland-Ross, M.-H. Qiu, I.~H.
  Gotlib, P.~M. Fuller, E.~Arrigoni, and J.~Lu, ``Identification of a direct
  gaba ergic pallidocortical pathway in rodents,'' \emph{European Journal of
  Neuroscience}, vol.~41, no.~6, pp. 748--759, 2015.

\bibitem{gast2021role}
R.~Gast, R.~Gong, H.~Schmidt, H.~G. Meijer, and T.~R. Kn{\"o}sche, ``On the
  role of arkypallidal and prototypical neurons for phase transitions in the
  external pallidum,'' \emph{Journal of neuroscience}, vol.~41, no.~31, pp.
  6673--6683, 2021.

\bibitem{cui2021striatal}
Q.~Cui, X.~Du, I.~Y. Chang, A.~Pamukcu, V.~Lilascharoen, B.~L. Berceau,
  D.~Garc{\'\i}a, D.~Hong, U.~Chon, A.~Narayanan \emph{et~al.}, ``Striatal
  direct pathway targets npas1+ pallidal neurons,'' \emph{Journal of
  Neuroscience}, vol.~41, no.~18, pp. 3966--3987, 2021.

\bibitem{labouesse2023non}
M.~Labouesse, A.~Torres-Herraez, M.~Chohan, J.~Villarin, J.~Greenwald, X.~Sun,
  M.~Zahran, A.~Tang, S.~Lam, J.~Veenstra-VanderWeele \emph{et~al.}, ``A
  non-canonical striatopallidal “go” pathway that supports motor control,''
  2023.

\bibitem{mizutani2017substance}
K.~Mizutani, S.~Takahashi, S.~Okamoto, F.~Karube, and F.~Fujiyama, ``Substance
  p effects exclusively on prototypic neurons in mouse globus pallidus,''
  \emph{Brain Structure and Function}, vol. 222, no.~9, pp. 4089--4110, 2017.

\bibitem{karube2019motor}
F.~Karube, S.~Takahashi, K.~Kobayashi, and F.~Fujiyama, ``Motor cortex can
  directly drive the globus pallidus neurons in a projection neuron
  type-dependent manner in the rat,'' \emph{Elife}, vol.~8, p. e49511, 2019.

\bibitem{nevado2014effective}
A.~J. Nevado-Holgado, N.~Mallet, P.~J. Magill, and R.~Bogacz, ``Effective
  connectivity of the subthalamic nucleus--globus pallidus network during
  parkinsonian oscillations,'' \emph{The Journal of physiology}, vol. 592,
  no.~7, pp. 1429--1455, 2014.

\bibitem{calabresi2014direct}
P.~Calabresi, B.~Picconi, A.~Tozzi, V.~Ghiglieri, and M.~Di~Filippo, ``Direct
  and indirect pathways of basal ganglia: a critical reappraisal,''
  \emph{Nature neuroscience}, vol.~17, no.~8, pp. 1022--1030, 2014.

\bibitem{kita2007globus}
H.~Kita, ``Globus pallidus external segment,'' \emph{Progress in brain
  research}, vol. 160, pp. 111--133, 2007.

\bibitem{plenz1999}
D.~Plenz and S.~T. Kital, ``A basal ganglia pacemaker formed by the subthalamic
  nucleus and external globus pallidus,'' \emph{Nature}, vol. 400, no. 6745,
  pp. 677--682, 1999.

\bibitem{terman2002}
D.~Terman, J.~E. Rubin, A.~Yew, and C.~Wilson, ``Activity patterns in a model
  for the subthalamopallidal network of the basal ganglia,'' \emph{Journal of
  Neuroscience}, vol.~22, no.~7, pp. 2963--2976, 2002.

\bibitem{lilascharoen2021divergent}
V.~Lilascharoen, E.~H.-J. Wang, N.~Do, S.~C. Pate, A.~N. Tran, C.~D. Yoon,
  J.-H. Choi, X.-Y. Wang, H.~Pribiag, Y.-G. Park \emph{et~al.}, ``Divergent
  pallidal pathways underlying distinct parkinsonian behavioral deficits,''
  \emph{Nature neuroscience}, vol.~24, no.~4, pp. 504--515, 2021.

\bibitem{farries2023selective}
M.~A. Farries, T.~W. Faust, A.~Mohebi, and J.~D. Berke, ``Selective encoding of
  reward predictions and prediction errors by globus pallidus subpopulations,''
  \emph{Current Biology}, vol.~33, no.~19, pp. 4124--4135, 2023.

\bibitem{isett2023indirect}
B.~R. Isett, K.~P. Nguyen, J.~C. Schwenk, J.~R. Yurek, C.~N. Snyder, M.~V.
  Vounatsos, K.~A. Adegbesan, U.~Ziausyte, and A.~H. Gittis, ``The indirect
  pathway of the basal ganglia promotes transient punishment but not motor
  suppression,'' \emph{Neuron}, 2023.

\bibitem{qiu2016deep}
M.~Qiu, M.~C. Chen, J.~Wu, D.~Nelson, and J.~Lu, ``Deep brain stimulation in
  the globus pallidus externa promotes sleep,'' \emph{Neuroscience}, vol. 322,
  pp. 115--120, 2016.

\bibitem{vetrivelan2010role}
R.~Vetrivelan, M.-H. Qiu, C.~Chang, and J.~Lu, ``Role of basal ganglia in
  sleep--wake regulation: neural circuitry and clinical significance,''
  \emph{Frontiers in neuroanatomy}, vol.~4, p. 145, 2010.

\bibitem{lazarus2013role}
M.~Lazarus, J.-F. Chen, Y.~Urade, and Z.-L. Huang, ``Role of the basal ganglia
  in the control of sleep and wakefulness,'' \emph{Current opinion in
  neurobiology}, vol.~23, no.~5, pp. 780--785, 2013.

\bibitem{stephenson2016basal}
M.~Stephenson-Jones, K.~Yu, S.~Ahrens, J.~M. Tucciarone, A.~N. van Huijstee,
  L.~A. Mejia, M.~A. Penzo, L.-H. Tai, L.~Wilbrecht, and B.~Li, ``A basal
  ganglia circuit for evaluating action outcomes,'' \emph{Nature}, vol. 539,
  no. 7628, pp. 289--293, 2016.

\bibitem{wallace2017genetically}
M.~L. Wallace, A.~Saunders, K.~W. Huang, A.~C. Philson, M.~Goldman, E.~Z.
  Macosko, S.~A. McCarroll, and B.~L. Sabatini, ``Genetically distinct parallel
  pathways in the entopeduncular nucleus for limbic and sensorimotor output of
  the basal ganglia,'' \emph{Neuron}, vol.~94, no.~1, pp. 138--152, 2017.

\bibitem{mallet2008parkinsonian}
N.~Mallet, A.~Pogosyan, L.~F. M{\'a}rton, J.~P. Bolam, P.~Brown, and P.~J.
  Magill, ``Parkinsonian beta oscillations in the external globus pallidus and
  their relationship with subthalamic nucleus activity,'' \emph{Journal of
  neuroscience}, vol.~28, no.~52, pp. 14\,245--14\,258, 2008.

\bibitem{crompe2020globus}
B.~d.~l. Crompe, A.~Aristieta, A.~Leblois, S.~Elsherbiny, T.~Boraud, and N.~P.
  Mallet, ``The globus pallidus orchestrates abnormal network dynamics in a
  model of parkinsonism,'' \emph{Nature communications}, vol.~11, no.~1, p.
  1570, 2020.

\bibitem{starr2008pallidal}
P.~A. Starr, G.~A. Kang, S.~Heath, S.~Shimamoto, and R.~S. Turner, ``Pallidal
  neuronal discharge in huntington's disease: support for selective loss of
  striatal cells originating the indirect pathway,'' \emph{Experimental
  neurology}, vol. 211, no.~1, pp. 227--233, 2008.

\bibitem{beste2015behavioral}
C.~Beste, M.~M{\"u}ckschel, S.~Elben, C.~J~Hartmann, C.~C~McIntyre, C.~Saft,
  J.~Vesper, A.~Schnitzler, and L.~Wojtecki, ``Behavioral and
  neurophysiological evidence for the enhancement of cognitive control under
  dorsal pallidal deep brain stimulation in huntington’s disease,''
  \emph{Brain Structure and Function}, vol. 220, pp. 2441--2448, 2015.

\bibitem{deng2021progression}
Y.~Deng, H.~Wang, M.~Joni, R.~Sekhri, and A.~Reiner, ``Progression of basal
  ganglia pathology in heterozygous q175 knock-in huntington's disease mice,''
  \emph{Journal of Comparative Neurology}, vol. 529, no.~7, pp. 1327--1371,
  2021.

\bibitem{baron2011multi}
M.~S. Baron, K.~D. Chaniary, A.~C. Rice, and S.~M. Shapiro, ``Multi-neuronal
  recordings in the basal ganglia in normal and dystonic rats,''
  \emph{Frontiers in systems neuroscience}, vol.~5, p.~67, 2011.

\bibitem{nambu2011reduced}
A.~Nambu, S.~Chiken, P.~Shashidharan, H.~Nishibayashi, M.~Ogura, K.~Kakishita,
  S.~Tanaka, Y.~Tachibana, H.~Kita, and T.~Itakura, ``Reduced pallidal output
  causes dystonia,'' \emph{Frontiers in Systems Neuroscience}, vol.~5, p.~89,
  2011.

\bibitem{starr2005spontaneous}
P.~A. Starr, G.~M. Rau, V.~Davis, W.~J. Marks~Jr, J.~L. Ostrem, D.~Simmons,
  N.~Lindsey, and R.~S. Turner, ``Spontaneous pallidal neuronal activity in
  human dystonia: comparison with parkinson’s disease and normal macaque,''
  \emph{Journal of neurophysiology}, vol.~93, no.~6, pp. 3165--3176, 2005.

\bibitem{chiken2008cortically}
S.~Chiken, P.~Shashidharan, and A.~Nambu, ``Cortically evoked long-lasting
  inhibition of pallidal neurons in a transgenic mouse model of dystonia,''
  \emph{Journal of Neuroscience}, vol.~28, no.~51, pp. 13\,967--13\,977, 2008.

\bibitem{hegeman2016external}
D.~J. Hegeman, E.~S. Hong, V.~M. Hern{\'a}ndez, and C.~S. Chan, ``The external
  globus pallidus: progress and perspectives,'' \emph{European Journal of
  Neuroscience}, vol.~43, no.~10, pp. 1239--1265, 2016.

\bibitem{yoshida2016two}
A.~Yoshida and M.~Tanaka, ``Two types of neurons in the primate globus pallidus
  external segment play distinct roles in antisaccade generation,''
  \emph{Cerebral Cortex}, vol.~26, no.~3, pp. 1187--1199, 2016.

\bibitem{gu2020globus}
B.-M. Gu, R.~Schmidt, and J.~D. Berke, ``Globus pallidus dynamics reveal covert
  strategies for behavioral inhibition,'' \emph{Elife}, vol.~9, p. e57215,
  2020.

\bibitem{arkadir2004independent}
D.~Arkadir, G.~Morris, E.~Vaadia, and H.~Bergman, ``Independent coding of
  movement direction and reward prediction by single pallidal neurons,''
  \emph{Journal of Neuroscience}, vol.~24, no.~45, pp. 10\,047--10\,056, 2004.

\bibitem{cui2021dissociable}
Q.~Cui, A.~Pamukcu, S.~Cherian, I.~Y. Chang, B.~L. Berceau, H.~S. Xenias, M.~H.
  Higgs, S.~Rajamanickam, Y.~Chen, X.~Du \emph{et~al.}, ``Dissociable roles of
  pallidal neuron subtypes in regulating motor patterns,'' \emph{Journal of
  Neuroscience}, vol.~41, no.~18, pp. 4036--4059, 2021.

\bibitem{mastro2017cell}
K.~J. Mastro, K.~T. Zitelli, A.~M. Willard, K.~H. Leblanc, A.~V. Kravitz, and
  A.~H. Gittis, ``Cell-specific pallidal intervention induces long-lasting
  motor recovery in dopamine-depleted mice,'' \emph{Nature neuroscience},
  vol.~20, no.~6, pp. 815--823, 2017.

\bibitem{georgopoulos1983relations}
A.~P. Georgopoulos, M.~R. DeLong, and M.~D. Crutcher, ``Relations between
  parameters of step-tracking movements and single cell discharge in the globus
  pallidus and subthalamic nucleus of the behaving monkey,'' \emph{Journal of
  Neuroscience}, vol.~3, no.~8, pp. 1586--1598, 1983.

\bibitem{mitchell1987primate}
S.~Mitchell, R.~Richardson, F.~Baker, and M.~DeLong, ``The primate globus
  pallidus: neuronal activity related to direction of movement,''
  \emph{Experimental Brain Research}, vol.~68, no.~3, pp. 491--505, 1987.

\bibitem{gage2010selective}
G.~J. Gage, C.~R. Stoetzner, A.~B. Wiltschko, and J.~D. Berke, ``Selective
  activation of striatal fast-spiking interneurons during choice execution,''
  \emph{Neuron}, vol.~67, no.~3, pp. 466--479, 2010.

\bibitem{turner1997pallidal}
R.~S. Turner and M.~E. Anderson, ``Pallidal discharge related to the kinematics
  of reaching movements in two dimensions,'' \emph{Journal of neurophysiology},
  vol.~77, no.~3, pp. 1051--1074, 1997.

\bibitem{turner2005context}
------, ``Context-dependent modulation of movement-related discharge in the
  primate globus pallidus,'' \emph{Journal of Neuroscience}, vol.~25, no.~11,
  pp. 2965--2976, 2005.

\bibitem{cui2013concurrent}
G.~Cui, S.~B. Jun, X.~Jin, M.~D. Pham, S.~S. Vogel, D.~M. Lovinger, and R.~M.
  Costa, ``Concurrent activation of striatal direct and indirect pathways
  during action initiation,'' \emph{Nature}, vol. 494, no. 7436, pp. 238--242,
  2013.

\bibitem{kravitz2010regulation}
A.~V. Kravitz, B.~S. Freeze, P.~R. Parker, K.~Kay, M.~T. Thwin, K.~Deisseroth,
  and A.~C. Kreitzer, ``Regulation of parkinsonian motor behaviours by
  optogenetic control of basal ganglia circuitry,'' \emph{Nature}, vol. 466,
  no. 7306, pp. 622--626, 2010.

\bibitem{schmidt2017pause}
R.~Schmidt and J.~D. Berke, ``A pause-then-cancel model of stopping: evidence
  from basal ganglia neurophysiology,'' \emph{Philosophical Transactions of the
  Royal Society B: Biological Sciences}, vol. 372, no. 1718, p. 20160202, 2017.

\bibitem{phillips2020}
R.~S. Phillips, I.~Rosner, A.~H. Gittis, and J.~E. Rubin, ``The effects of
  chloride dynamics on substantia nigra pars reticulata responses to pallidal
  and striatal inputs,'' \emph{Elife}, vol.~9, p. e55592, 2020.

\bibitem{alexander1986parallel}
G.~E. Alexander, M.~R. DeLong, and P.~L. Strick, ``Parallel organization of
  functionally segregated circuits linking basal ganglia and cortex,''
  \emph{Annual review of neuroscience}, vol.~9, no.~1, pp. 357--381, 1986.

\bibitem{delgado2023}
L.~Delgado-Zabalza, N.~P. Mallet, C.~Glangetas, G.~Dabee, M.~Garret,
  C.~Miguelez, and J.~Baufreton, ``Targeting parvalbumin-expressing neurons in
  the substantia nigra pars reticulata restores motor function in parkinsonian
  mice,'' \emph{Cell Reports}, vol.~42, no.~10, 2023.

\end{thebibliography}

\end{document}